\documentclass[structabstract]{aa}
\usepackage{graphicx}
\usepackage{longtable}
\usepackage{supertabular}
\usepackage{lscape}
\usepackage{hhline}
\usepackage{setspace}
\usepackage{threeparttable}
\usepackage{txfonts}
\begin{document}
\title{A study of the remarkable galaxy system AM\,546-324\\ (the core of Abell S0546)
\thanks{Based on observations made at the Gemini Observatory, under the identification number GS-2010B-Q-7.}}

\author{M. Fa\'{u}ndez-Abans\inst{1}
\and A.~C. Krabbe\inst{2}
\and M.~de Oliveira-Abans\inst{1,3}
\and P.~C. da Rocha-Poppe \inst{4,5}\\
I. Rodrigues \inst{2}
\and V.~A. Fernandes-Martin\inst{4,5}
\and I.~F. Fernandes\inst{4,5}
          }

\offprints{Max Fa\'{u}ndez-Abans; max@lna.br}

\institute{MCTI/Laborat\'{o}rio  Nacional  de  Astrof\'{\i}sica, Rua Estados Unidos, 154, Bairro das Na\c{c}\~{o}es, CEP 37.504-364, Itajub\'{a}, MG, Brazil \\
\email{mfaundez@lna.br, mabans@lna.br}
\and Universidade do Vale do Para\'{i}ba - UNIVAP. Av. Shishima Hifumi, 2911 - Urbanova
    CEP: 12244-000 - S\~{a}o Jos\'{e} dos Campos, SP, Brazil.
             \email{angela.krabbe@gmail.com, irapuan@univap.br}
\and UNIFEI, Instituto de Engenharia de Produ\c{c}\~{a}o e Gest\~{a}o, Av. BPS 1303 Pinheirinho, 37500-903 Itajub\'{a}, MG, Brazil
\and  UEFS, Departamento de F\'{i}sica, Av. Transnordestina, S/N, Novo Horizonte, Feira de Santana, BA, Brazil, CEP 44036-900
\and UEFS, Observat\'{o}rio Astron\^{o}mico Antares, Rua da Barra, 925, Jardim Cruzeiro, Feira de Santana, BA, Brazil, CEP 44015-430 \\
\email{paulopoppe@gmail.com, vmartin1963@gmail.com, irafbear@gmail.com}
}
   \date{Received 17 February 2012 / Accepted 7 May 2012}

% \abstract{}{}{}{}{} 
% 5 {} token are mandatory
 
\abstract
{}
{We report first results of an investigation of the tidally disturbed galaxy system AM\,546-324, whose two principal
galaxies 2MFGC 04711 and AM\,0546-324 (NED02) were previously classified as interacting doubles. This system was selected to study the interaction of ellipticals in a moderately dense environment. We provide spectral characteristics of the system and present an observational study of the interaction effects on the morphology, 
kinematics, and stellar population of these galaxies.
}
{The study is based on long-slit spectrophotometric data in the range of $\sim$ 4500-8000 $\AA$ obtained with the Gemini Multi-Object Spetrograph at Gemini South (GMOS-S). We have used the stellar population synthesis code STARLIGHT to investigate the star formation history of these galaxies. The Gemini/GMOS-S direct r-G0303 broad band pointing image was used to enhance and study fine morphological structures. The main absorption lines in the spectra were used to determine the radial velocity.
}
{Along the whole long-slit signal, the spectra of the Shadowy galaxy (discovered by us), 2MFGC 04711, and AM\,0546-324 (NED02) resemble that of an early-type galaxy. We  estimated redshifts of z= 0.0696, z= 0.0693 and z= 0.0718, corresponding to heliocentric velocities of 20\,141 km s$^{-1}$, 20\,057 km s$^{-1}$, and 20\,754 km s$^{-1}$ for the Shadowy galaxy, 2MFGC 04711 and AM\,0546-324 (NED02), respectively. The central regions of 2MFGC 04711 and AM\,0546-324 (NED02) are completely dominated by an old stellar population of \mbox{$2\times10^{9} <\rm t \leq 13\times10^{9}$ yr} and do not show any spatial variation in the contribution of the stellar-population components.}
{The observed rotation profile distribution of 2MFGC 04711 and AM\,0546-324 (NED02) can be adequately interpreted as an ongoing stage of interaction with the Shadowy galaxy as the center of the local gravitational potential-well of the system. The three galaxies are all early-type. The extended and smooth distribution of the material in the Shadowy galaxy is a good laboratory to study  direct observational signatures of tidal friction in action.
}
\keywords{galaxies: general -- galaxies: interacting group -- individual: AM\,0546-324 and 2MFGC 04711 --
              galaxies: spectroscopy -- galaxies: stellar synthesis}

\titlerunning{The galaxy pair AM\,546-324}
\authorrunning{Fa\'undez-Abans et al.}
\maketitle

%
%________________________________________________________________

\section{Introduction}

Galaxy interactions and mergers are fundamentally important in the formation and evolution of galaxies.
Hierarchical models of galaxy formation and various observational evidence suggest that elliptical galaxies are, like disk galaxies, embedded in massive dark-matter halos. Lenticular and elliptical galaxies, called early-type galaxies have been thought to be the end point of galaxy evolution. These systems have shown uniform red optical colors and display a tight red sequence in optical color-magnitude diagrams (e.g. Baldry et al. \cite{b2004}). Their color separation from star-forming galaxies is thought to be due to a lack of fuel for star formation, which must have been consumed, destroyed or removed on a reasonably short timescale (e.g. Faber et al. \cite{f2007}). In addition, numerical simulations have shown that the global characteristics of the binary merger remnants of two equal-mass spiral galaxies, called major mergers, resemble those of early-type galaxies (Toomre \& Toomre \cite{tt1972}; Hernquist \& Barnes \cite{hb1991}; Barnes \cite{b1992}; Mihos et al. \cite{m1995}; Springel \cite{s2000}; Naab \& Burkert \cite{nb2003}; Bournaud, Jog \& Combes \cite{bjc2005}). Remnants with properties similar to early-type objects can also be recovered through a multiple minor merger process, the total accreted mass of which is at least half of the initial mass of the main progenitor (Weil \& Hernquist \cite{wh1994}, \cite{wh1996}; Bournaud, Jog \& Combes \cite{bjc2007}). This scenario of early-type formation through accretion and merging of bodies would fit well within the frame of the hierarchical assembly of galaxies provided by cold dark matter cosmology.

Interactions between early-type galaxies are less spectacular than those observed in spiral galaxies. While impressive tidal tails, plumes, bridges, and shells are observed in tidally disturbed spirals,  the effects  of the interaction are  less easily recognized in elliptical galaxies, since they have little gas and dust, and are dominated essentially by old stellar populations. Evidence for recent merger-driven star formation  (Rogers et al.  \cite{r09})  and morphological  disturbances such as shells, ripples, and rings have been observed in early-type galaxies (Kaviraj et al. \cite{kv10} and Wenderoth et al. \cite{wen2011}) .
 
The peculiar Ring Galaxies (pRGs) show a wide variety of ring and bulge morphologies and were classified by  Fa\'{u}ndez-Abans \& de Oliveira-Abans (\cite{foa98a}) into five families, following the general behavior of galaxy-ring structures. From these categories eight morphological subdivisions are highlighted.  One of these morphological subdivisions is a basic structure called Solitaire. The pRG Solitaire is described as an object with the bulge on the ring, or very close to it, resembling a one-diamond finger ring (single knotted ring). In these objects, the ring generally looks smooth and thinner on the opposite side of the bulge (as archetypes \object{FM\,188-15/NED02}, \object{AM\,0436-472/NED01)}, \object{ESO 202-IG45/NED01} and \object{ESO 303-IG11/NED01}). Although the statistics are as yet poor, the Solitaire type is probably produced by the interaction between elliptical-like galaxies and/or gas-poor S0 galaxies with an elliptical companion. In a forthcoming paper, a list of Solitaire-type pRGs and a preliminary study and statistics will be presented.

There are no reports of Solitaires in early stages of formation in the literature yet; so a few pairs of galaxies were selected as early-stage candidates (see one of them in Wenderoth et al. \cite{wen2011}). Even though one of the selected candidates,  \mbox{AM\,546-324}, originally extracted from Arp \& Madore's catalog (Arp \& Madore \cite{am1977}, \cite{am1986}; category 2, interacting doubles), seems to be  morphologically different from an expected Solitaire in the early stage, it is remarkable enough to be studied as an almost isolated ``spherical/elliptical and S0 interacting objects" in centrally sparse clusters of galaxies.

In this paper, we report new results for the tidally disturbed galaxy system \object{AM\,546-324} based on data obtained from long-slit spectrophotometric observations at Gemini Observatory, in Chile. Values of \mbox{$H_{\rm o}$ = 70 km s$^{-1}{\rm Mpc}^{-1}$}, $\Omega_{matter} = 0.27$ and $\Omega_{vaccum} = 0.73$ have been adopted throughout this work (Freedman et al. \cite{f2001}; Astier et al. \cite{a2006}, and Spergel et al. \cite{s2003}).

\section{AM\,0546-324 review}

The existing information on this object comprises: (1) the Arp-Madore catalog (Arp \& Madore \cite{am1986}) referred to it as  ``Category 2: interacting doubles", which are objects consisting of two galaxies that, by their apparent magnitude and spacing, appear to be associated; (2) the redshifts of 165 Southern rich Abell cluster of galaxies by Quintana \& Ram\'{i}rez (\cite{qr1995}), in which the galaxy system \object{AM\,546-324} is a member of Abell S0546; (3) the 2MASS-selected flat galaxy catalog by Mitronova et al. (\cite{mit2004}); (4) the catalog of near-infrared properties of LEDA galaxies using the full-resolution images from the DENIS survey (Paturel et al. \cite{pat2005}); and (5)  the entry from the 6dFGS-NVSS data by Mauch \& Sadler (\cite{ms2007}). In Abell et al. (\cite{aco1989}), S0546 is quoted as irregular following the cluster classification in Abell's (\cite{a1965}) system, with 23 cluster members. Using the data quoted in Abell et al. (\cite{aco1989}) in a magnitude-redshift relation for the Abell southern clusters, and using the S0546 distance class $m_{10}=$5 result in an approximate radial velocity of cz= 20\,893 km\,s$^{-1}$, which agrees with our results (using the non-relativistic velocity formula, see  Table~\ref{table1}).

\begin{figure}
\centering
\resizebox{80 mm}{!}{\includegraphics[clip]{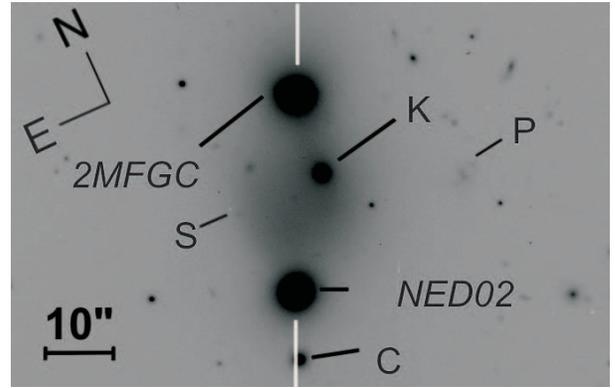}}
\caption{System AM\,0546-324. Optical 5-min exposure GMOS-S image in the r-G0303 
filter, enhanced by a median filter kernel of 300$\times$300 pixels 
(see Fa\'{u}ndez-Abans \& de Oliveira-Abans \cite{foa98b} for details on the method). The slit position \mbox{PA = 157$\degr$} is also displayed as two short white lines to preserve the image of the objects. The letters stand for: 
K, ``the Knot" (Quintana \& Ram\'{i}rez \cite{qr1995}); S, the Shadowy galaxy; C, a companion galaxy; and 
P, a probable Polar Ring galaxy.}
\label{fig_01}
\end{figure}

Figure~\ref{fig_01} displays the \object{AM\,546-324} system in a 5 minute-exposure GMOS-S pointing 
image in the r-G0303 filter (effective wavelength of 6300 $\AA$). Table~\ref{table1} displays the
 new velocity and z values, together with some early information on the principal members of 
 \object{AM\.0546-324}, Table~\ref{table2} displays some information on relevant objects in and around the system.

% -------------------------------- Figure 1

% ----------------------------> Table 1

\begin{table*}
\caption{ Basic properties of the principal galaxies of the system AM\,0546-324}
\label{table1}
\centering
\begin{tabular}{lllll}
\hline \hline
Parameter & 2MFGC 04711 & Shadowy galaxy & AM\,0546-324 (NED02) & Ref.\\
\hline
%              &          &     &      \\
 R.A. (2000)  & 05 48 34.1 & 05 48 34.7 & 05 48 35.1 & this work \\
Dec. (2000) & \hspace{-0.1cm}-32 39 30.9  & \hspace{-0.1cm}-32 39 46.2 & \hspace{-0.1cm}-32 40 01.0  & this work \\
Morphological classification & Elliptical & Cd? & Elliptical & this work \\
$z$ & 0.0693 & 0.0696  & 0.0718 & this work \\
$V$(km\,s$^{-1}$) ({\it a}) & 20\,057 $\pm$10 & 20\,141 $\pm$10 & 20\,754 $\pm$10& this work \\
$V$(km\,s$^{-1}$) ({\it b}) & 20\,793 $\pm$10 & 20\,880 $\pm$10 & 21\,526 $\pm$10& this work \\
\hline
$z$  & 0.0692 &  &  0.0721 & NED ({\it q})\\
$V$ (km\,s$^{-1}$) & 20\,749 $\pm$40 &  & 21\,615 $\pm$26 & NED ({\it q})\\
Magnitude & 15.0 R &  & & NED\\
Other designations & 2MASX J05483415-3239306 & & 2MASX J05483518-3240006 & NED\\
\hline
Distance (Mpc) & 292.6 & 293.8 & 303.0 & this work \\
Distance (Mpc) & 293.0 &  & 305.0 & NED \\
$\sigma_{v}$ (km/s) &  312 & 365 & 197 & this work \\
Mass (lower limit) & 1.63$\times10^{11}$ M$_{\sun}$ & 5.24$\times10^{11}$ M$_{\sun}$ & 1.60$\times10^{11}$ M$_{\sun}$ & this work \\
U-shaped base ({\it r}) & 2\farcs0  & & 1\farcs54 &this work \\
Major axis (lower limit) ({\it r}) & 9\farcs42 & 21\farcs73 & 8\farcs53 &this work \\
J - H & 0.404 & & 0.732 & NED \\
H - K & 0.142 & & \hspace{-0.2cm}$-$0.050 & NED \\
J - K & 0.546 & & 0.641 & NED \\
\hline
\end{tabular}
\begin{tablenotes}
\item[a]{Note: {\it (a)}: extracted for high velocities (Lang \cite{lang1999}), see also Lindengren \& Dravins (\cite{ld2003}); {\it (b)}: non-relativistic velocity using the standard formula (Lang \cite{lang1999}); {\it (r)}: measurements on the Gemini/GMOS-S {\it r-G0303}-filter; ({\it q}): original data from Quintana \& Ram\'{i}rez (\cite{qr1995}).}
\end{tablenotes}
\end{table*}

% ----------------------------> Table 2

\begin{table*}
\caption{Relevant objects in and around system AM\,0546-324.}
\label{table2}
\centering
\begin{tabular}{lllllll}
\hline \hline
Object & R.A.(2000) & Dec.(2000) &$z_{\rm abs}$ & $V$(km\,s$^{-1}$) & Distance (Mpc) & Ref.\\
\hline
J054832.5-323954.1 & 05 48 32.5 & -32 39 54.1 &        &        & & Polar Ring? this work \\
J054834.2-323944.5 & 05 48 34.2 & -32 39 44.5 & 0.0698 & 20\,923 & 298.9 & The Knot ({\it a}) \\
J054835.4-324011.5 & 05 48 35.4 & -32 40 11.5 & 0.0685 & 20\,550 $\pm$40 & 293.6& C, this work ({\it b})\\
\hline
\end{tabular}
\begin{tablenotes}
\item[a]{Note: {\it (a)} Quintana \& Ram\'{i}rez (\cite{qr1995}); {\it (b)} the compact companion C close to \object{AM\,0546-324 (NED02)}.}
\end{tablenotes}
\end{table*}

\section{Observations and data reduction}

The spectroscopic observations were performed with the 8.1-m Gemini South telescope (Chile) (ID program GS-2010B-Q-7). We used the GMOS-S spectrograph in long-slit mode (Hook et al. \cite{h04})\footnote{A 
description of the instrument can be found at http://www.gemini.edu/sciops/instruments/gmos.}. 
The R400+G5325 grating was centered at \mbox{6\,250 $\AA$} and used with a long-slit 1.5 arcsec x 375 arcsec. The data were binned by 2 in the spatial dimension and 2 in the spectral dimension,  producing a spectral resolution of $\sim$5.1$\AA$ FWHM, sampled at 0.68 $\AA$ pix$^{-1}$. The seeing throughout the observations was 0\farcs54 and the binned pixel scale was 0\farcs145 pix$^{-1}$. The wavelength range was $\sim$ 4500-8000 $\AA$. The 
spectrophotometric standard star H\,600 was observed using the same experimental set up. The long-slit 
spectra were taken at one position angle on the sky, PA = 157$\degr$, to encompass the three objects in one shot, 
and it almost crossed the center of each object.

The standard Gemini-IRAF routines were used to carry out bias subtraction, flat-fielding, and cosmic ray subtraction. The data were then wavelength calibrated with an accuracy $\leqslant$ 0.3 $\AA$. The binned 2-D spectra were then flux-calibrated using the photometric standard star  H\,600. The 2-D spectra were then extracted into 1-D spectra, which were sky-subtracted and binned in the spatial dimension. We have cross-correlated our observed spectra with three galaxy and star templates with good signal-to-noise. These results were checked with the composite absorption-line template ``fabtemp97" distributed by RVSAO\footnote{The RVSAO IRAF (Radial Velocity Package for IRAF) external package was developed at the Smithsonian Astrophysical Observatory. Full documentation of this software, including numerous examples of its use, in on-line at http://tdc-www.harvard.edu/iraf/rvsao/.}/IRAF external package. We adopted the redshift value from the best highest correlated coefficient template. 

The spectral apertures were extracted with the APALL/IRAF package and three methods: (1) the standard IRAF procedure; (2) overlapping a shifted sample with steps \mbox{$< 5\arcsec$}, which causes oversampling; (3) and the event-covering method, for which we used the aperture step as a mapping event process (Wong \& Chiu \cite{wc1987}\footnote{The event-covering method is defined as a strategy of selecting a subset of statistically independent events in a set of variable-pairs, regardless of their statistical independence.}; see also Schafer \cite{s1997} and Wu \& Barbara \cite{wb2002}), which also causes oversampling. The idea of the last two procedures was to use the aperture size as a filter to detect kinematical structures in the long-slit velocity map. The results of the first two procedures have been used in this work. The third method was mainly used when the original data were either corrupted or incomplete. Its results were not different from those of the second method because of the data completeness.

%__________________________________________________________________

\section{Analysis and results}

\subsection{The field around \object{AM\,0546-324}}

\begin{figure}[h]
\begin{center}$
\begin{array}{cc}
\resizebox{\hsize}{!}{\includegraphics[]{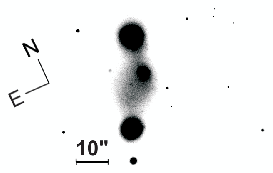}}\\
\resizebox{\hsize}{!}{\includegraphics[]{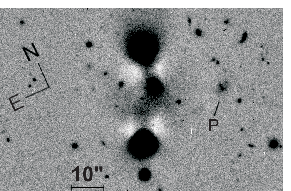}}
\end{array}$
\end{center}
\caption{From top to bottom: (a) median filtering with a 100$\times$100-pixel kernel after the original image subtraction, (b)  same as first panel for a 500$\times$500-pixel kernel. The clear patches are artifacts of the method, but the fine structures are preserved. The candidate for a  Polar Ring, which has been discovered in this work, is marked by the letter ``P".}
\label{fig_02}
\end{figure}

As can be seen in Fig. \ref{fig_01}, this system of galaxies is seen almost edge-on, with five prominent objects in the field. According to the estimated distances, they  may be  physically associated. In  Fig. \ref{fig_01},  from NW to SE, these objects are (1) the almost spherical galaxy 2MFGC 04711; (2) the galaxy ``Knot", K, as labeled by  Quintana \& Ram\'{i}rez (\cite{qr1995}); (3) the object that we have named the Shadowy galaxy (hereafter the S galaxy), whose center is enhanced in the figure; (4) AM\,0546-324 (NED02) (hereafter NED02), which is almost spherical; and (5) the  compact anonymous spherical galaxy ``C". Another relevant object to the SW in the field (better seen in Fig. \ref{fig_02}, bottom panel) is a probable Polar Ring galaxy, named here ``P", whose redshift value is not known yet.

\begin{figure}[h]
\begin{center}$
\begin{array}{cc}
\resizebox{\hsize}{!}{\includegraphics[]{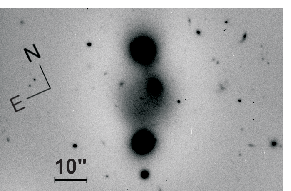}}\\
\resizebox{90 mm}{!}{\includegraphics[]{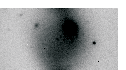}}
\end{array}$
\end{center}
\caption{From top to bottom: (a) resulting image after using a low-pass filter (see text), 
(b) same as first panel for a Gauss filter, zooming on the S galaxy.}
\label{fig_03}
\end{figure}

To extract as much information as possible from the GMOS-S r-G0303 pointing image, we used different spatial filtering to find fine morphological structures in the frame. The top panel of Fig.~\ref{fig_02} displays the result of applying a median filter kernel of 100$\times$100 pixels, where the S galaxy appears elongated, its center and a few rims having also been enhanced. A few thin filaments in the K object, with one of them pointing to the center of the S galaxy, have also been enhanced. Furthermore, the galaxies K, \object{2MFGC 04711}, and NED02 appear to be slightly deformed by the tidal interaction. The lower panel of Fig.~\ref{fig_02} displays a median filter kernel of 500$\times$500 pixels where the deformation of the main objects and some faint dwarf structures apparently bound to this system have been enhanced. The rims and the center of  S galaxy are still evident. The upper panel of Fig.~\ref{fig_03} shows the center and the rims of the S galaxy, after a low-pass filter was used. It highlights the deformation of the elliptical galaxies and enhances a few notable dwarf satellites around this system. The bottom panel of Fig.~\ref{fig_03} is a zoom on the S galaxy to better illustrate its center and a few very faint rims, after using a Gaussian filter.

To determine the central ellipticity of the galaxies, we used the {\it ELLIPSE} STSDAS-task\footnote{Space Telescope Science Data Analysis Software Package}, which fits elliptical isophotes to galaxy direct images. Then we created a 2-D noiseless model image using the {\it BMODEL} STSDAS-task built from the results of the isophotal analysis. Table~\ref{table3} lists rough estimates of the ``non perturbed" elliptical-class section and the whole major axis-diameter in kpc for the quoted galaxies.

% -------------------------------- Figure 2

% -------------------------------- Figure 3

% ----------------------------> Table 3

\begin{table}
\caption{Ellipticity and major axis diameter.}
\label{table3}
\centering
\begin{tabular}{lcc}
\hline \hline
Object & ``Bulge-like" section & Major diameter \\
       & (elliptical class) & (kpc) \\
\hline
2MFGC 04711 & E1 & 13.7 \\
Shadowy & \hspace{+0.2cm}E3/4:: & 31.6: \\
NED02 & E0 & 12.4 \\
The Knot & \hspace{+0.1cm}E1: & \hspace{+0.1cm}9.5 \\
C companion &  E1 & \hspace{+0.1cm}3.6 \\
\hline
\end{tabular}
\end{table}

\subsection{The spectra and kinematics}

\begin{figure}[h]
\begin{center}$
\begin{array}{cc}
\resizebox{\hsize}{!}{\includegraphics[]{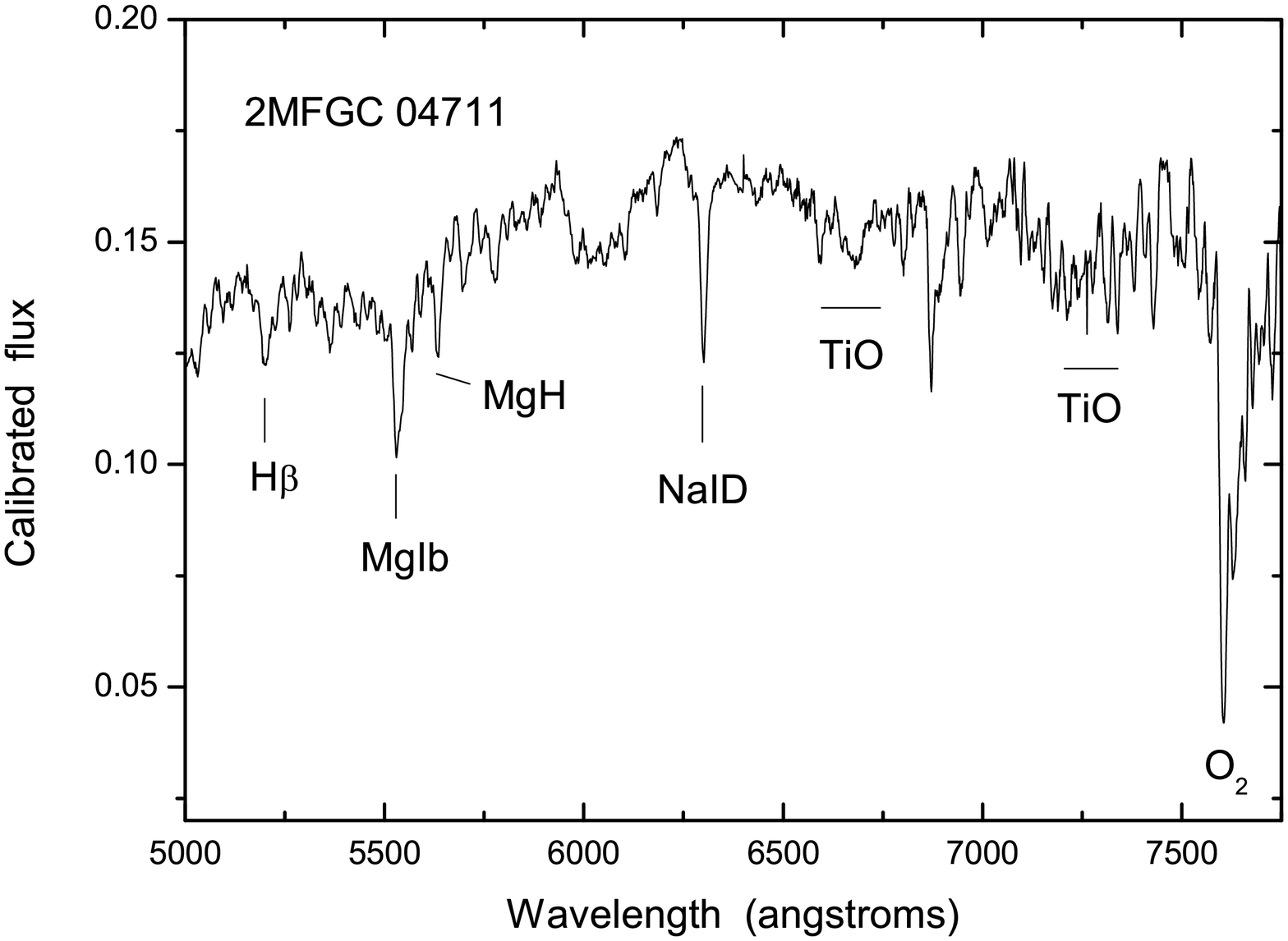}}\\
\resizebox{\hsize}{!}{\includegraphics[]{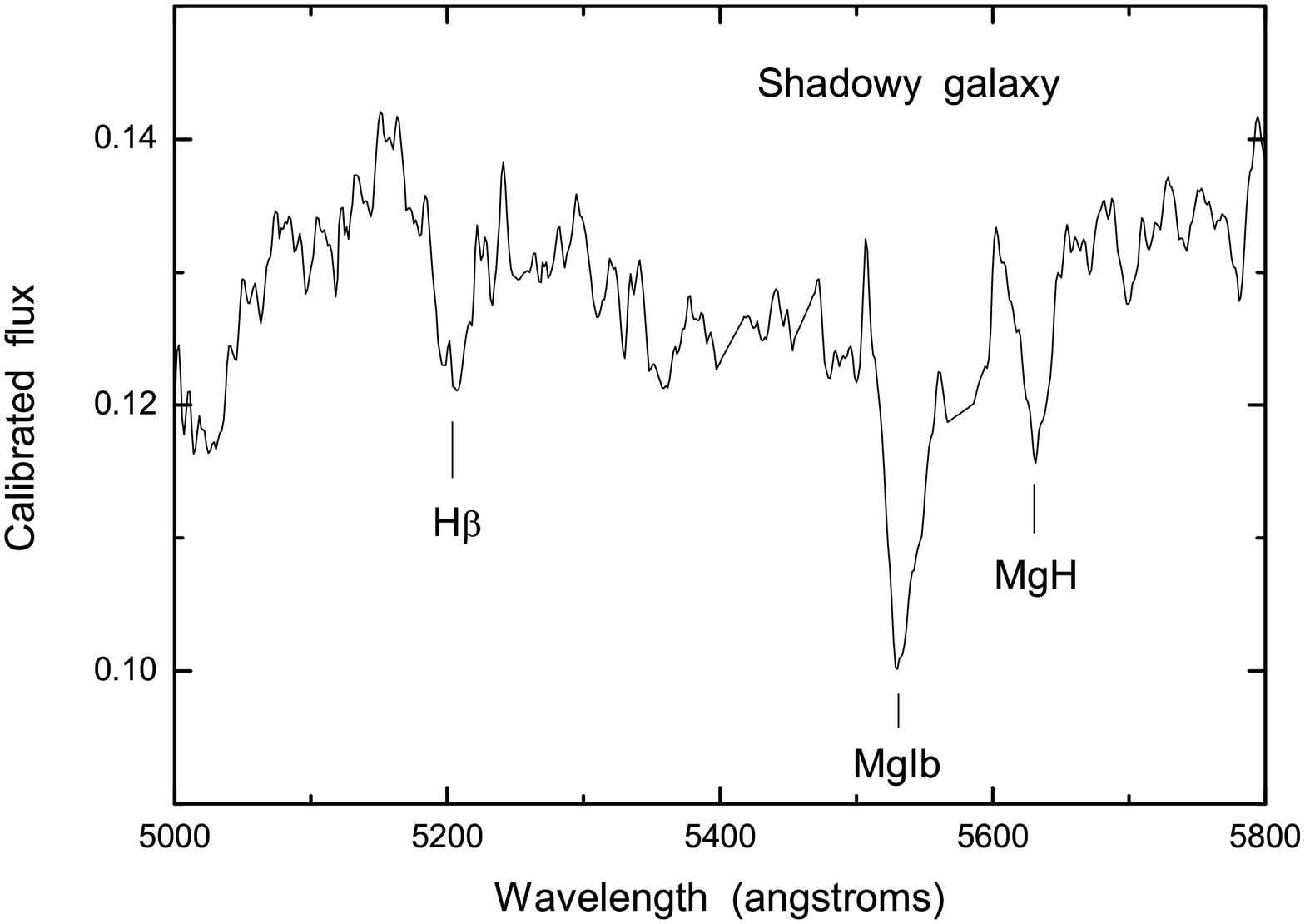}}\\
\resizebox{\hsize}{!}{\includegraphics[]{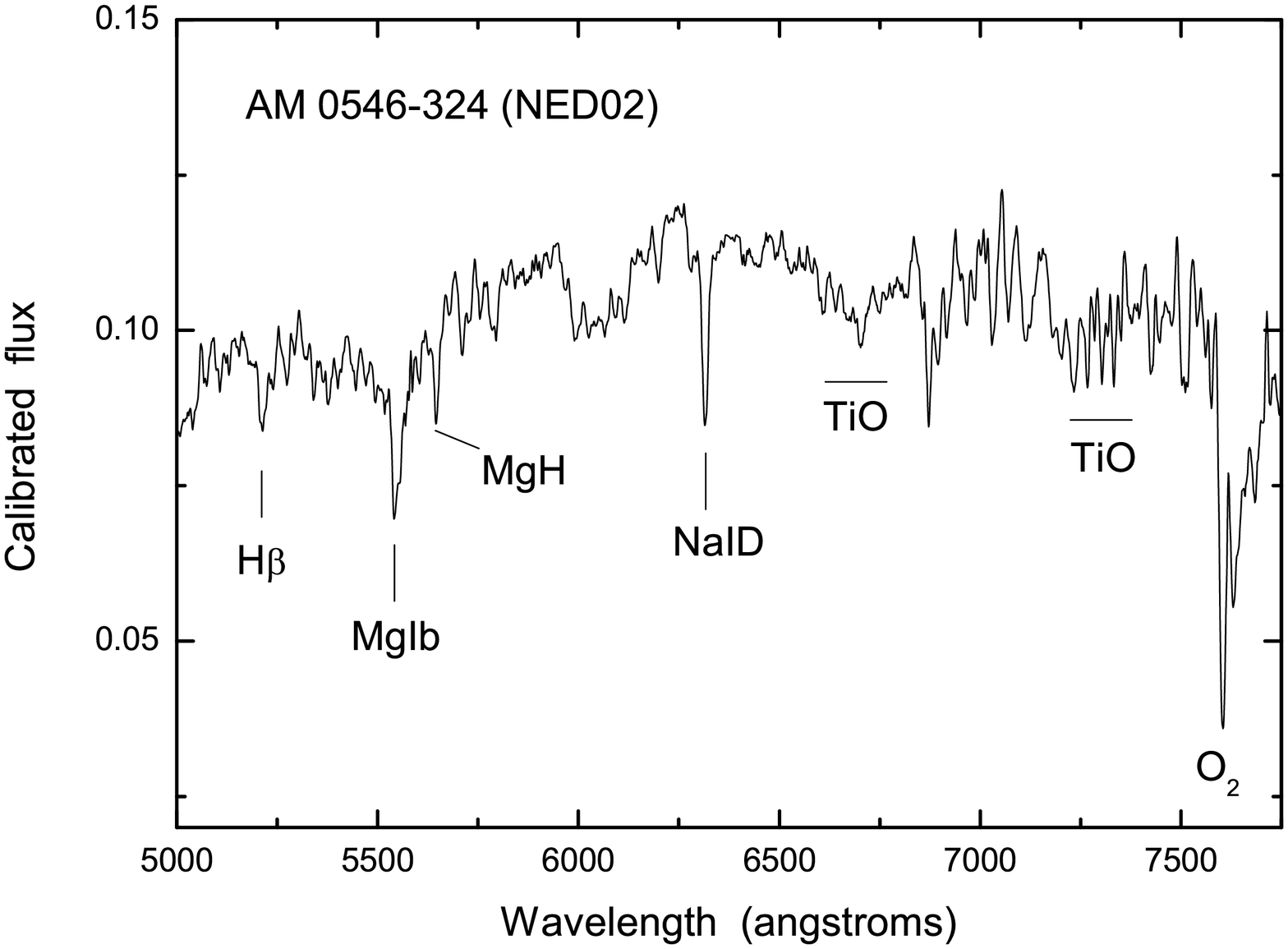}}
\end{array}$
\end{center}
\caption{Optical nuclear absorption features of \object{2MFGC 04711}, the Shadowy galaxy and \object{AM\,0546-324 (NED02)}, in the upper, middle and lower panels, respectively. 
The spectra are displayed  in units of $\times10^{-16}$ erg sec$^{-1}$cm$^{-2}\AA^{-1}$.}
\label{fig_04}
\end{figure}

We report the first dedicated long-slit spectroscopic results for the three main galaxies of the AM\,0546-324 system. A sample of the spectra of these galaxies are shown in Fig.~\ref{fig_04}.  The $\lambda\lambda5000-7750$~\AA \,spectral section of 2MFGC 04711 and NED02 are displayed, while the S galaxy's is displayed only in the $\lambda\lambda5000-5800$~\AA \,spectral section. They show the main stellar absorption features identified in the nuclear region: H$\beta$, MgIb$\lambda5174$, MgH$\lambda5269$, NaID$\lambda5892$, the TiO bands $\lambda\lambda6250,7060$, and the O$_2$ atmospheric band. The column density and positions of these lines were determined by fitting a Gaussian to the observed profile. This procedure was reviewed using the RVSAO/{\it XCSAO} package. The spectra in the regions $\lambda\lambda7750-8000$~\AA  \,(for 2MFGC 04711 and NED02) and $\lambda\lambda5800-8000$~\AA \,(for the S galaxy) were not taken into account because of the high noise. The characteristics of the spectra are discussed below. 

% -------------------------------- Figure 4

% -------------------------------- Figure 5

\begin{figure}[h]
\begin{center}$
\begin{array}{cc}
\resizebox{\hsize}{!}{\includegraphics[]{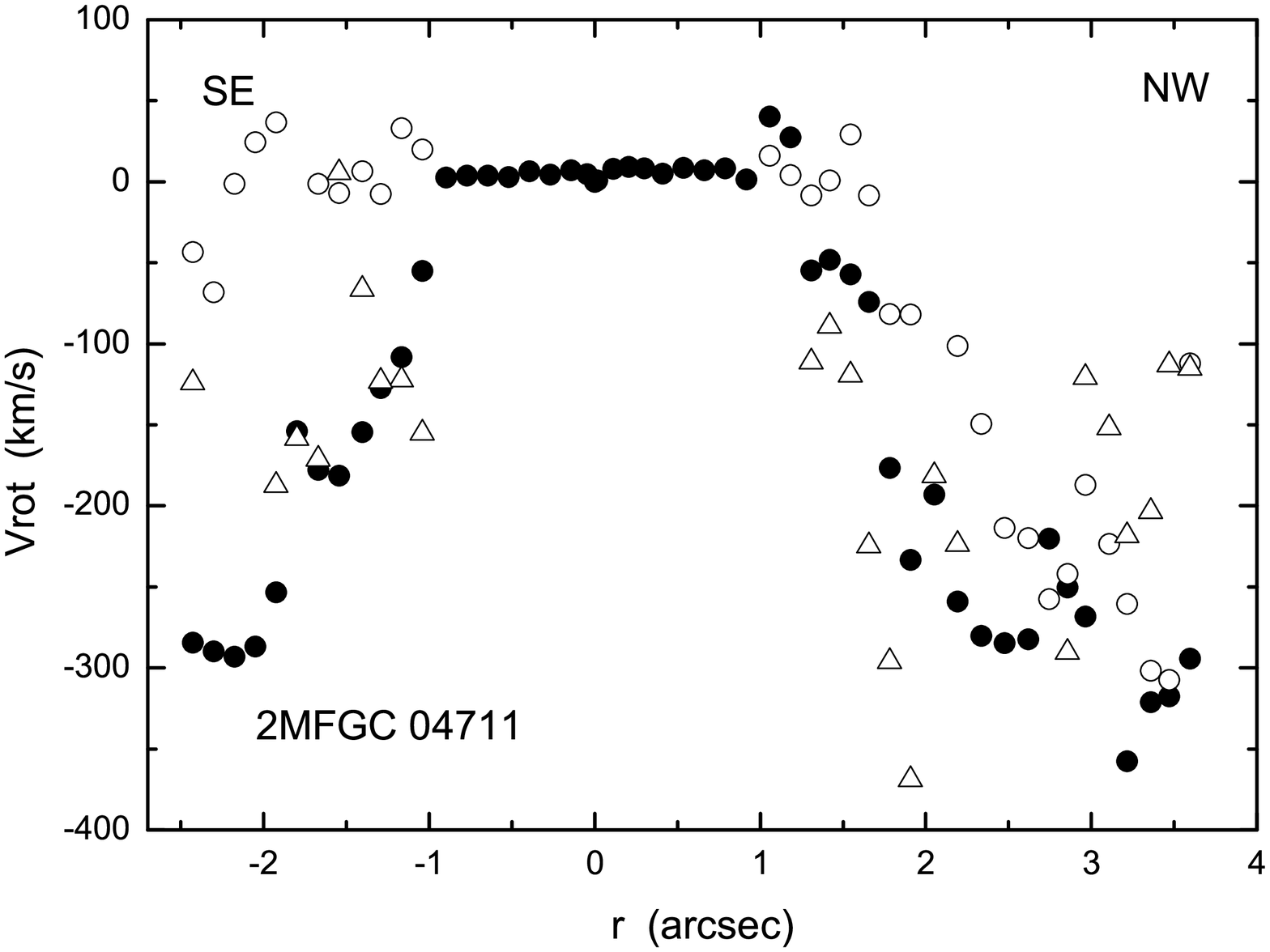}}\\
\resizebox{\hsize}{!}{\includegraphics[]{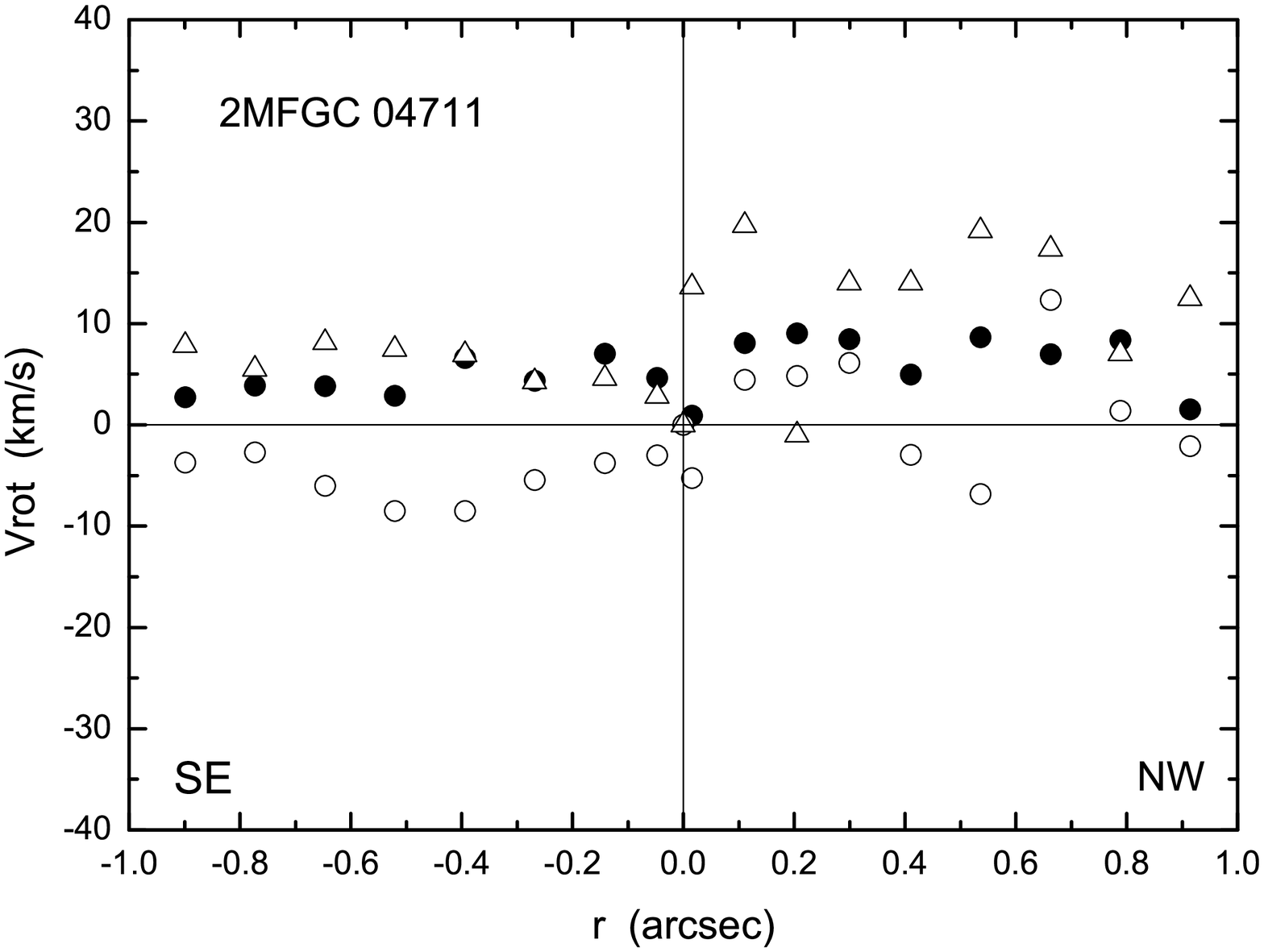}}
\end{array}$
\end{center}
\caption{(a) upper panel: the U-shaped rotation profile of \mbox{2MFGC 04711} along the nucleus and bulge in the total observed long-slit $-$3\arcsec \,to $+$4\arcsec distribution (filled circles are data from NaID lines, open circles from MgIb and open triangles stand for H$\beta$ lines); (b) lower panel: same as first panel, for the central $-$1\arcsec \,to $+$1\arcsec core features, but using a shifted aperture sample with steps \mbox{$< 2\arcsec$} according to method (2) cited in item \S3.}
\label{fig_05}
\end{figure}

% ---------------------------------Figure 6

\begin{figure}[h]
\begin{center}$
\begin{array}{cc}
\resizebox{\hsize}{!}{\includegraphics[]{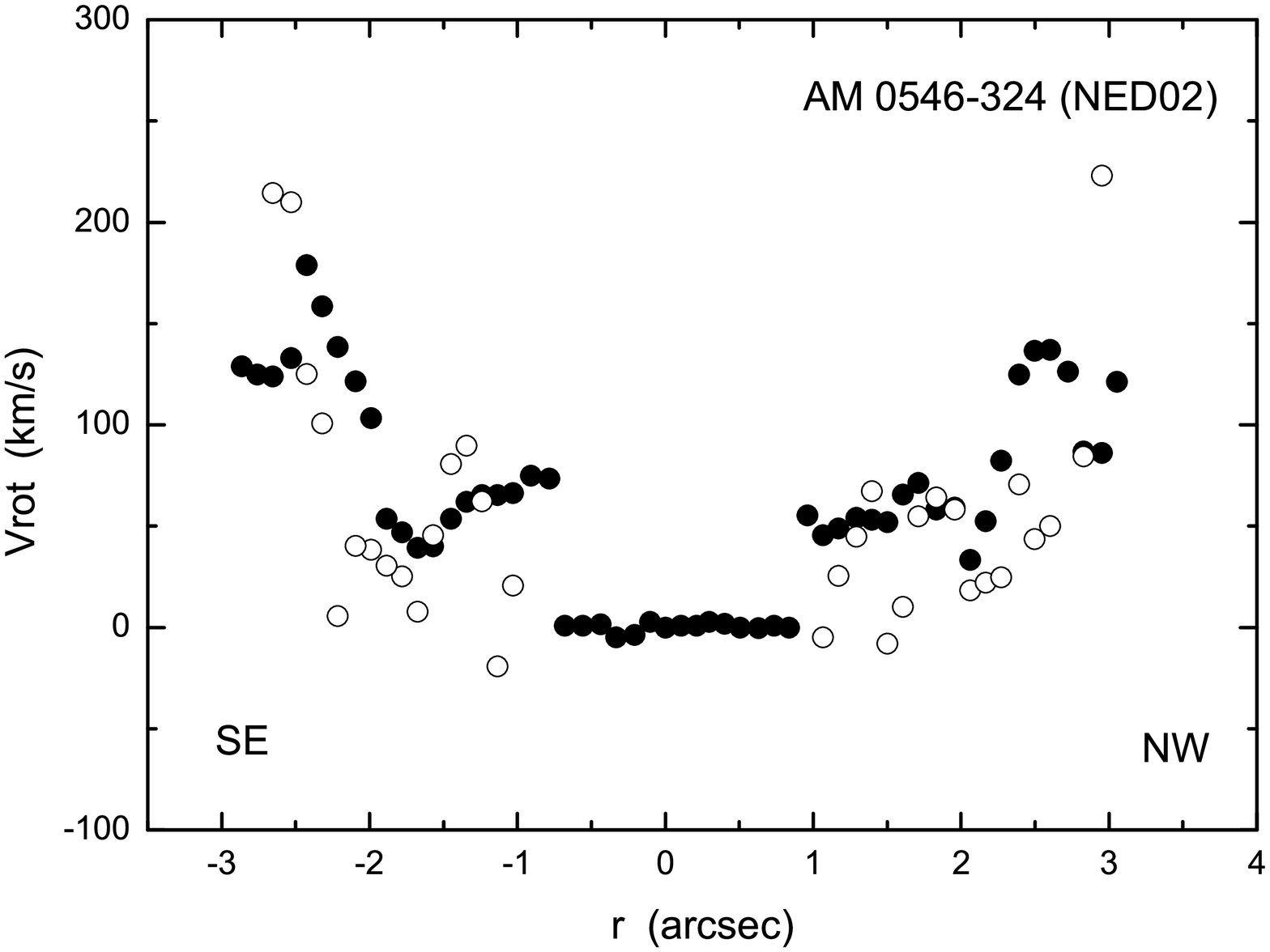}}\\
\resizebox{\hsize}{!}{\includegraphics[]{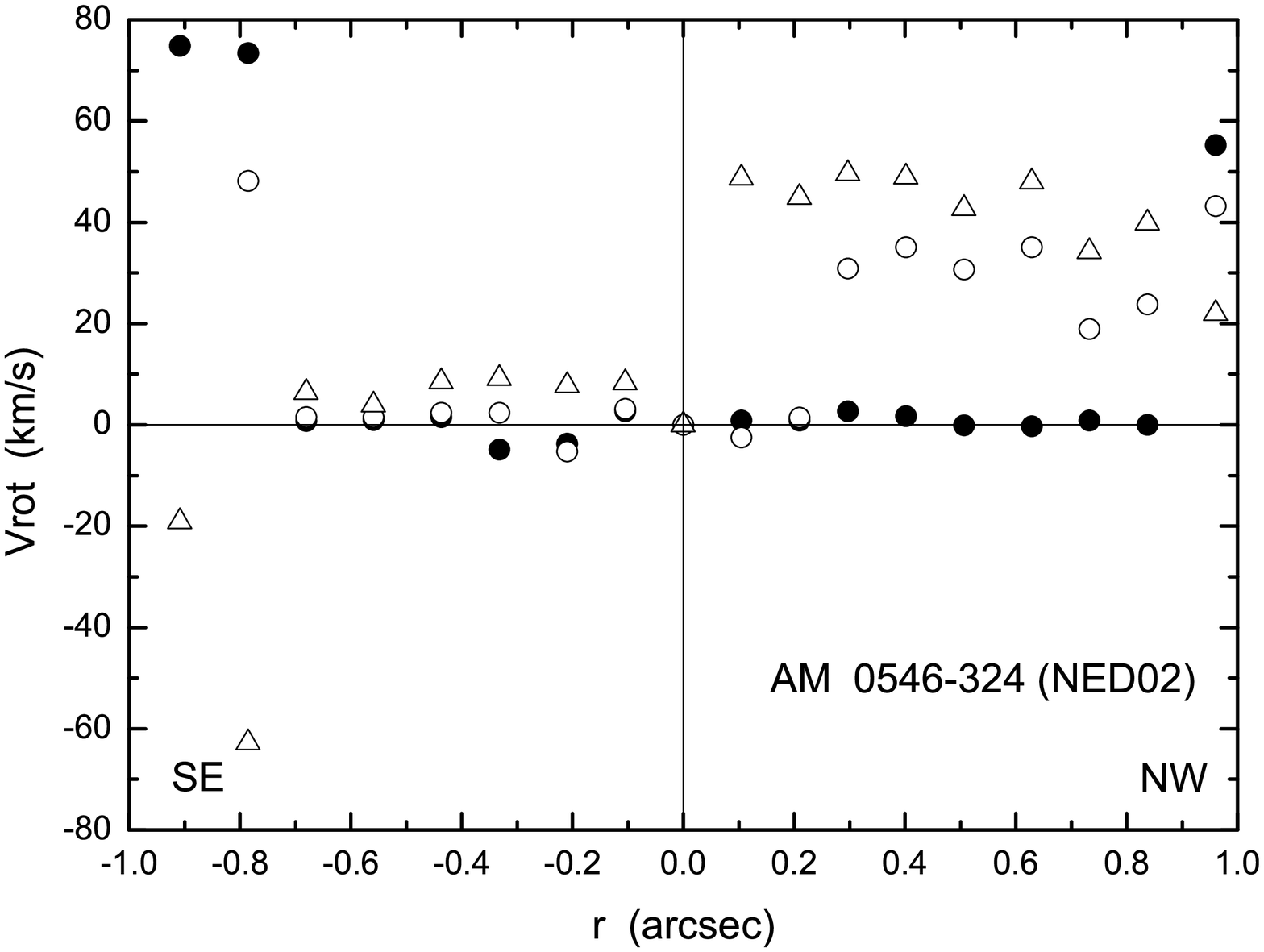}}
\end{array}$
\end{center}
\caption{From top to bottom: (a) the U-shaped rotation profile of \object{AM\,0546-324 (NED02)} along the nucleus and bulge in the total observed slit length $-$3\arcsec \,to $+$4\arcsec (filled circles are data from NaID lines, open circles from MgIb); (b) same as first panel, for the central $-$1\arcsec to $+$1\arcsec core features, but using a shifted aperture sample with steps \mbox{$< 2\arcsec$} according to method (2) cited in item \S3.}
\label{fig_06}
\end{figure}

\subsubsection{\object{2MFGC 04711}}

No emission lines were detected along the slit. The spectral profile resembles that of an early-type galaxy. Using the NaID, MgIb, and H$\beta$ absorption lines, we cross-correlated our observed spectra with templates with good signal-to-noise and these results were checked with the composite absorption-line template ``fabtemp97" distributed by RVSAO. We derived z=0.0693 and a heliocentric radial velocity of $V$= 20\,057 $\pm$10 km\,s$^{-1}$ . The velocity dispersion $\sigma_{v}$ values were estimated by cross-correlation with K and M star templates, with statistical error in $\sigma_{v}$ in the range of 10\%$-$15\% (XCSAO/RVSAO, also tested with XCOR/STSDAS). The dynamical masses were estimated based on a virial relation, the effective radius and the velocity dispersion of each galaxy. The derived effective radius and dynamical mass fit within $\le1\sigma$ with the size-mass relation presented by Bezanson et al. (\cite{beza2011}), predicts values close to those derived here for each galaxy. The calculated distance and dynamical mass are 292.6 Mpc and 1.63$\times10^{11}$ M$_{\sun}$, respectively. Figure~\ref{fig_05} displays the \object{2MFGC 04711} distribution of radial velocities measured from the H$\beta$, MgIb and NaID absorption lines, along the slit-section (upper panel) and central region (lower panel). The errors of the individual velocity measurements do not exceed 10 km\,s$^{-1}$ in the central region and increase to 20$-$40 km\,s$^{-1}$ on its periphery.

\subsubsection{AM\,0546-324 (NED02)}

NED02 also shows an early-type spectral profile. We estimated z=0.0718 and a heliocentric radial velocity of $V$= 20\,754 $\pm$10 km\,s$^{-1}$. The calculated distance and dynamical mass are 303 Mpc and 1.60$\times10^{11}$ M$_{\sun}$, respectively. Figure~\ref{fig_06} displays the NED02 distribution of radial velocities measured from the MgIb and NaID absorption lines, along the slit-section (upper panel) and central region (lower panel). For clarity, the H$\beta$ data are not shown in the first panel because they are mostly coincident with the NaID points at r $>\pm$1\arcsec \,. The errors of the individual velocity measurements do not exceed 10 km\,s$^{-1}$ in the central region and increase to 20$-$40 km\,s$^{-1}$ on its periphery.

\subsubsection{The Shadowy galaxy}

We estimated z=0.0696 and a heliocentric radial velocity of $V$= 20\,141 $\pm$10 km\,s$^{-1}$, from an integrated central spectral-section of 3\farcs5. On the other hand, the calculated velocities for the NW and SE sections of the galaxy are 20\,193 $\pm$20 km\,s$^{-1}$ (NW) and 20\,081 $\pm$20 km\,s$^{-1}$ (SE), respectively (integrated NW$-$SE spectral-section of 6\farcs1). The calculated distance and  lower limit of the dynamical mass are 293.8 Mpc and 5.24$\times10^{11}$ M$_{\sun}$, respectively.

The distribution of the radial velocities of 2MFGC 04711 and NED02 shows the U-shaped rotation profile (first panel of Figs.~\ref{fig_05} and~\ref{fig_06}). This shape has been reported in studies of interacting binary-disturbed elliptical galaxies (see Borne \cite{b1990}; Borne \& Hoessel \cite{bh1985},\cite{bh1988}; Bender et al. \cite{bpn1991}; Madejsky \cite{mad1991} and Madejsky et al. \cite{madall1991}). The U-shaped profiles are common in strongly interacting elliptical galaxies, and the physical interpretation, given by Borne et al. (\cite{borne1994}) is that there is a tidal coupling between the orbit of the companion and the resonant prograde rotating stars in the kinematically disturbed galaxy (Borne \cite{bor1988}; Borne \& Hoessel \cite{bh1988}; Bacells et al. \cite{bacells1989}). The coupling of NED02 and 2MFGC 047111 with the S galaxy and the U-shaped rotation profile of these galaxies are thus a direct observational signature of tidal friction in action within this system, in agreement with the physical interpretation of Borne et al. (\cite{borne1994}). Based on the merging times of simulations performed for a low-mass galaxy falling on to a massive elliptical by Leeuwin \& Combes (\cite{lc1997}) and adopting for the S galaxy a r$_{max}$ = 30 kpc, our rough estimate for the decay times for 2MFGC 04711, NED02 and C galaxy are 4$\times10^{8}$ \,yr, 3.6$\times10^{8}$ \,yr, and 5.6$\times10^{8}$ \,yr,
respectively. With the suitably simple model reported by Leeuwin \& Combes (\cite{lc1997}) the amount of friction should accelerate in the decay time, which could be the scenario for the S satellite with a decay time shorter than 1.0$\times10^{9}$ \,yrs. 

The errors of the individual velocity measurements do not exceed 10 km\,s$^{-1}$ in the central region of the galaxies and increase to 20$-$40 km\,s$^{-1}$ at their periphery. There is a significant dispersion in the radial velocity distribution of both 2MFGC and NED02 (see first panels of Figs.~\ref{fig_05} and~\ref{fig_06}, respectively). The velocity spread is $\pm$40$-$110 km\,s$^{-1}$ around $\pm$1\arcsec$-$3\arcsec.

\subsection{Stellar population synthesis}

The detailed study of star formation in tidally perturbed galaxies provides important 
information not only on the age distribution of the stellar population, but also helps to 
better understand  several aspects related to the interacting process and its effects on the properties 
of the individual galaxies and their evolution.

To investigate the star formation history of NED02 and 2MFGC, we used the stellar population synthesis code {\sc STARLIGHT} (Cid Fernandes et al. \cite{cid04,cid05}; Asari et al.\cite{asari07}).
This code has been extensively discussed in Cid Fernandes et al. (\cite{cid04,cid05}) and is built upon computational techniques originally developed for empirical population synthesis with additional ingredients from evolutionary synthesis models. This method was also used by Krabbe et al. (\cite{krabbe2011}) and has been successful in describing the stellar population in interacting galaxies.

The code fits an observed spectrum $O_{\lambda}$ with a combination of $N_{\star}$ single stellar populations (SSPs) from the Bruzual \& Charlot (\cite{bruzual03}) models. These models are based on a high-resolution library of observed stellar spectra, which allows for detailed spectral evolution of the SSPs at a resolution of 3\,\AA\, across the wavelength range of 3\,200-9\,500 $\AA$ with a wide range of metallicities.  We used the Padova (1994) tracks as recommended by Bruzual \& Charlot (\cite{bruzual03}), with the initial mass function of Salpeter (\cite{salpeter55}) between 0.1 and 100 $M_{\sun}$. Extinction is modeled by {\sc STARLIGHT} as due to foreground dust, using the Large Magellanic Cloud average reddening law of Gordon et al. (\cite{gordon03}) 
with  R$_V$= 3.1, and parametrized by the V-band extinction  A$_V$. The SSPs used in this work cover 15 ages namely, t = 0.001\,, 0.003\,, 0.005\,, 0.01\,, 0.025\,, 0.04\,, 0.1\,,  0.3\,, 0.6\,, 0.9\,, 1.4\,, 2.5\,, 5\,, 11\,, and 13 Gyr, as well as three metallicities, Z = 0.2 Z$_{\sun}$, 1 Z$_{\sun}$, and 2.5 Z$_{\sun}$, adding to  45 SSP components. The fitting is carried out using  a simulated annealing plus Metropolis scheme, with bad pixel regions excluded from the analysis.

Prior to the modeling, the SSPs models were convolved to the same resolution of the observed spectra;
the observed spectra were shifted to their rest-frame,  corrected for foreground Galactic reddening of $E(B-V)=0.036$ mag taken from Schlegel et al. (\cite{sc98}) and normalized to $\lambda\,5870\,$\AA. The error in $O_{\lambda}$ considered in the fitting was the continuum rms with a $S/N  \ge 10$, where $S/N$ is the signal-to-noise  ratio per $\AA$ in the region around $\lambda_{0}=5870\,\AA$. In addition, the fitting was performed only in spectra with absorption lines.

\begin{figure}
\centering
\includegraphics*[angle=-90,width=\columnwidth]{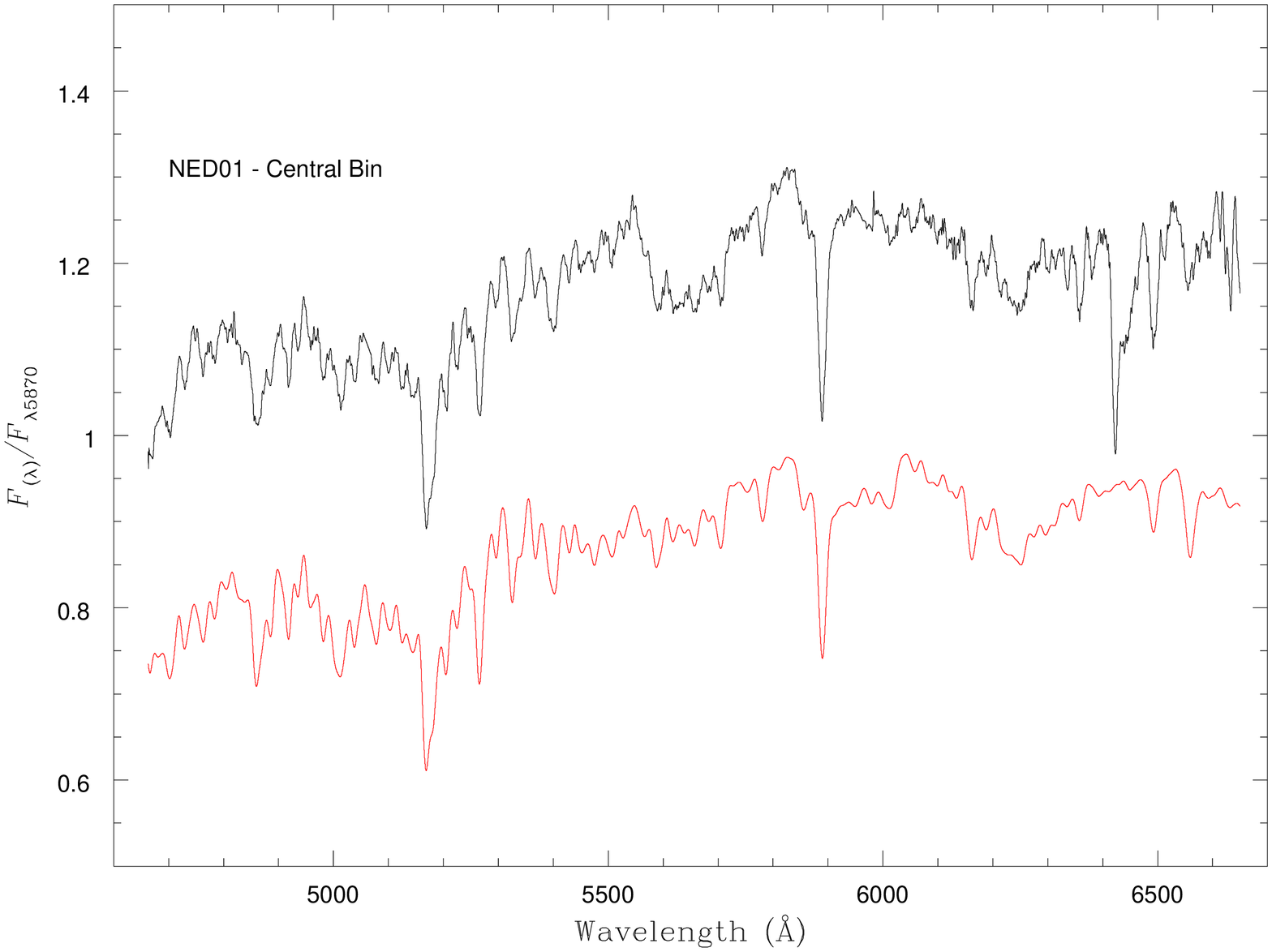}						     
\caption{Stellar population synthesis for 2MFGC 04711. Central bin observed spectrum corrected for reddening (in black, shifted up by a constant)  and the synthesized spectrum, in red.}
\label{fig_07}
\end{figure}

\begin{figure}
\centering
\includegraphics*[angle=-90, width=\columnwidth]{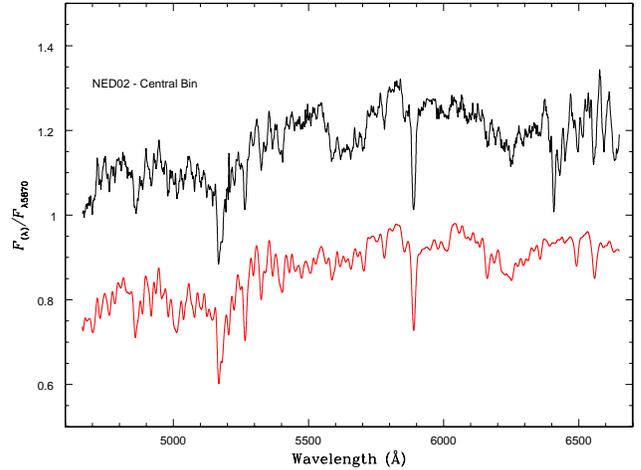}						     
\caption{Stellar population synthesis for NED02. Central bin observed spectrum corrected for reddening (in black, shifted up by a constant)  and the synthesized spectrum, in red.}
\label{fig_08}
\end{figure}

Figures \ref{fig_07} and \ref{fig_08}  show an example of the observed spectrum corrected by reddening and the
model stellar population spectrum for 2MFGC 04711 and AM\,0546-324 (NED02) galaxies, respectively. The results of the synthesis  to the very central region are summarized in Table \ref{synt_table} for the individual spatial bins in each galaxy, stated as the perceptual contribution of each base element to the flux  at $\lambda\, 5\,870$\, \AA. Following the prescription of Cid Fernandes et al. (\cite{cid05}), we have defined a condensed population vector by binning the stellar populations according to the flux contributions into young, \mbox{$x_{\rm Y}$ ($\rm t \leq 5\times10^{7}$ yr)};
intermediate-age,  \mbox{$x_{\rm I}$ ($ 5\times10^{7} <\rm t \leq 2\times10^{9}$ yr)}; and 
old, \mbox{$x_{\rm O}$ ( $2\times10^{9} <\rm t \leq 13\times10^{9}$ yr)} components. The same bins were used to
represent the mass components of the population vector $m_{\rm Y}$, $m_{\rm I}$, and
$m_{\rm O}$). The metallicity (Z), one important parameter to characterize the stellar population content,
is weighted by light fraction. The quality of the fitting result is measured by the parameters 
$\chi^{2}$ and  $adev$. The latter gives the perceptual mean deviation $|O_{\lambda} - M_{\lambda}|/O_{\lambda}$ over all fitted pixels, where $O_{\lambda}$ and $M_{\lambda}$ are the observed and model spectra, respectively. 

The spatial variation in the contribution of the stellar population components for 2MFGC. NED02 is completely dominated by an old stellar population.

\begin{table*}
\caption{Stellar-population synthesis results}
\label{synt_table}
\begin{tabular}{lrrrrrrrrrr}
 \\
\noalign{\smallskip}
\hline
\hline
\noalign{\smallskip}
Pos. &  \multicolumn{1}{c}{$ x_{\rm Y}$} &  \multicolumn{1}{c}{$x_{\rm I}$} &  \multicolumn{1}{c}{$x_{\rm O}$}
 & \multicolumn{1}{c}{$ m_{\rm Y}$}& \multicolumn{1}{c}{$ m_{\rm I}$} &
 \multicolumn{1}{c}{$m_{\rm O}$}& 
  \multicolumn{1}{c}{$Z_{\star}$[1]} &
 \multicolumn{1}{c}{$ \chi^{2}$} & 
 \multicolumn{1}{c}{$\rm adev$} & \multicolumn{1}{c}{$\rm A_{v}$}
 \\
 
\multicolumn{1}{c}{(arcsec)}& \multicolumn{1}{c}{(per cent)}  &
 \multicolumn{1}{c}{(per cent)} &  \multicolumn{1}{c}{(per cent)}
 & \multicolumn{1}{c}{(per cent)} & 
 \multicolumn{1}{c}{(per cent)}
   &
 \multicolumn{1}{c}{(per cent)}& 
  \multicolumn{1}{c}{} &
 \multicolumn{1}{c}{} & 
 \multicolumn{1}{c}{(mag)}
 \\
\hline
\noalign{\smallskip}
\noalign{\smallskip}
\multicolumn{10}{c}{2MFGC 04711}\\
\noalign{\smallskip}
\hline
\noalign{\smallskip}
-1.04   &   0.0  &  0.0  &   100.0   &   0.0  &   0.0  &   100.0  &   0.020  &	1.9  &  2.24  &  0.59 \\
-0.90   &   0.0  &  0.0  &   100.0   &   0.0  &   0.0  &   100.0  &   0.027  &	1.9  &  1.24  &  0.07 \\
-0.77   &   0.0  &  0.0  &   100.0   &   0.0  &   0.0  &   100.0  &   0.026  &	2.0  &  1.24  &  0.07 \\
-0.65   &   0.0  &  0.0  &   100.0   &   0.0  &   0.0  &   100.0  &   0.027  &	1.9  &  1.25  &  0.07 \\
-0.52   &   0.0  &  0.0  &   100.0   &   0.0  &   0.0  &   100.0  &   0.028  &	2.0  &  1.21  &  0.07 \\
-0.39   &   0.0  &  0.0  &   100.0   &   0.0  &   0.0  &   100.0  &   0.027  &	2.0  &  1.24  &  0.07 \\
-0.27   &   0.0  &  0.0  &   100.0   &   0.0  &   0.0  &   100.0  &   0.026  &	1.9  &  1.24  &  0.07 \\
-0.14   &   0.0  &  0.0  &   100.0   &   0.0  &   0.0  &   100.0  &   0.028  &	2.0  &  1.22  &  0.07 \\
-0.05   &   0.0  &  0.0  &   100.0   &   0.0  &   0.0  &   100.0  &   0.027  &	2.0  &  1.23  &  0.07 \\
0.0	&   0.0  &  0.0  &   100.0   &   0.0  &   0.0  &   100.0  &   0.023  &	1.8  &  1.30  &  0.00 \\
0.02    &   0.0  &  0.0  &   100.0   &   0.0  &   0.0  &   100.0  &   0.027  &	2.0  &  1.22  &  0.07 \\
0.11    &   0.0  &  0.0  &   100.0   &   0.0  &   0.0  &   100.0  &   0.027  &	2.0  &  1.22  &  0.07 \\
0.20    &   0.0  &  0.0  &   100.0   &   0.0  &   0.0  &   100.0  &   0.027  &	2.0  &  1.26  &  0.07 \\
0.30    &   0.0  &  0.0  &   100.0   &   0.0  &   0.0  &   100.0  &   0.027  &	1.9  &  1.24  &  0.07 \\
0.41    &   0.0  &  0.0  &   100.0   &   0.0  &   0.0  &   100.0  &   0.027  &	1.9  &  1.23  &  0.07 \\
0.54    &   0.0  &  0.0  &   100.0   &   0.0  &   0.0  &   100.0  &   0.027  &	1.9  &  1.24  &  0.07 \\
0.66    &   0.0  &  0.0  &   100.0   &   0.0  &   0.0  &   100.0  &   0.027  &	1.9  &  1.24  &  0.07 \\
0.79    &   0.0  &  0.0  &   100.0   &   0.0  &   0.0  &   100.0  &   0.027  &	1.9  &  1.23  &  0.07 \\
0.91    &   0.0  &  0.0  &   100.0   &   0.0  &   0.0  &   100.0  &   0.026  &	2.0  &  1.25  &  0.07 \\
1.06    &   0.0  &  15.0 &   85.0    &   0.0  &   4.9  &    95.1  &   0.041  &	1.9  &  1.93  &  0.00 \\
1.18    &   0.0  &  21.0 &   79.0    &   0.0  &   5.6  &    94.4  &   0.034  &	1.8  &  2.02  &  0.00 \\
Integrated	  &   0.0  &   0.0  &   100.0  &   0.0  &   0.0  &   100.0  &   0.026  &   1.9  &  1.23  &  0.06 \\
\noalign{\smallskip}
\hline
\noalign{\smallskip}
\multicolumn{10}{c}{AM\,0546-324 (NED02)}\\
\noalign{\smallskip}
\hline
\noalign{\smallskip}
-0.68     &   0.0  &      0.0  &  100.0    &   0.0  &	  0.0  &   100.0  &   0.026  &   2.2  &  1.41  &  0.02 \\
-0.56     &   0.0  &      0.0  &  100.0    &   0.0  &	  0.0  &   100.0  &   0.026  &   2.2  &  1.43  &  0.02 \\
-0.44     &   0.0  &      0.0  &  100.0    &   0.0  &	  0.0  &   100.0  &   0.025  &   2.2  &  1.36  &  0.02 \\
-0.33     &   0.0  &      0.0  &  100.0    &   0.0  &	  0.0  &   100.0  &   0.027  &   2.2  &  1.43  &  0.02 \\
-0.21     &   0.0  &      0.0  &  100.0    &   0.0  &	  0.0  &   100.0  &   0.027  &   2.2  &  1.41  &  0.02 \\
-0.10     &   0.0  &      0.0  &  100.0    &   0.0  &	  0.0  &   100.0  &   0.026  &   2.2  &  1.39  &  0.02 \\
0.00  	  &   0.0  &      0.0  &  100.0    &   0.0  &	  0.0  &   100.0  &   0.025  &   2.2  &  1.24  &  0.02 \\
0.10      &   0.0  &	  0.0  &  100.0    &   0.0  &	  0.0  &   100.0  &   0.026  &   2.3  &  1.35  &  0.02 \\
0.21      &   0.0  &	  0.0  &  100.0    &   0.0  &	  0.0  &   100.0  &   0.026  &   2.2  &  1.34  &  0.02 \\
0.30      &   0.0  &	  0.0  &  100.0    &   0.0  &	  0.0  &   100.0  &   0.027  &   2.2  &  1.44  &  0.02 \\
0.40      &   0.0  &	  0.0  &  100.0    &   0.0  &	  0.0  &   100.0  &   0.026  &   2.2  &  1.43  &  0.02 \\
0.51      &   0.0  &	  0.0  &  100.0    &   0.0  &	  0.0  &   100.0  &   0.025  &   2.2  &  1.35  &  0.02 \\
0.63      &   0.0  &	  0.0  &  100.0    &   0.0  &	  0.0  &   100.0  &   0.025  &   2.2  &  1.42  &  0.02 \\
0.73      &   0.0  &	  0.0  &  100.0    &   0.0  &	  0.0  &   100.0  &   0.026  &   2.2  &  1.38  &  0.02 \\
0.84      &   0.0  &	  0.0  &  100.0    &   0.0  &	  0.0  &   100.0  &   0.024  &   2.4  &  1.63  &  0.17 \\
Integrated &   0.0  &   0.0  &   100.0  &   0.0  &   0.0  &   100.0  &   0.020  &   1.7  &  1.54  &  0.00 \\
\noalign{\smallskip}
\hline
\noalign{\smallskip}
\noalign{\smallskip}
\noalign{\smallskip}
\end{tabular}
\begin{minipage}[c]{18.0cm}
[1] Abundance by mass with Z$_{\sun}$=0.02 \\
\end{minipage}
\end{table*}

% ---------------------------------------------- Figure 07

% ---------------------------------------------- Figure 08

\section{Discussion}

\begin{figure}[h]
\centering
\resizebox{\hsize}{!}{\includegraphics[]{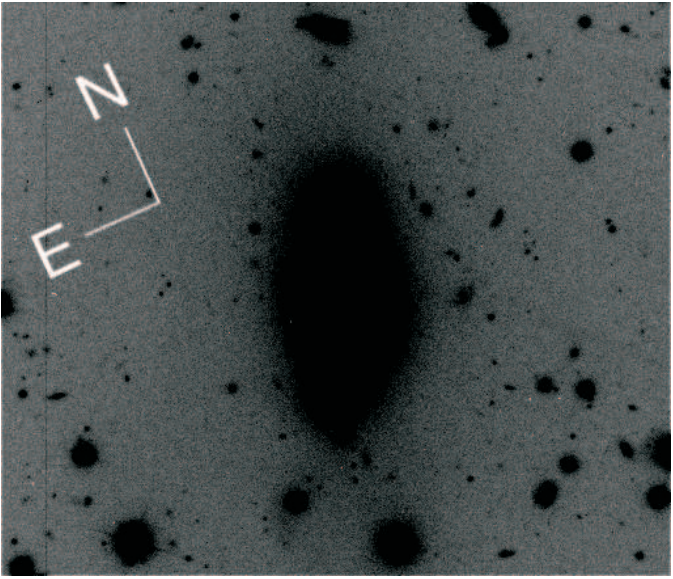}}
\caption{Reproduction of the image of AM\,0546-324 with different contrast to highlight the dwarf objects crowding the system.}
\label{fig_09}
\end{figure}

\begin{figure}[h]
\centering
\resizebox{\hsize}{!}{\includegraphics[]{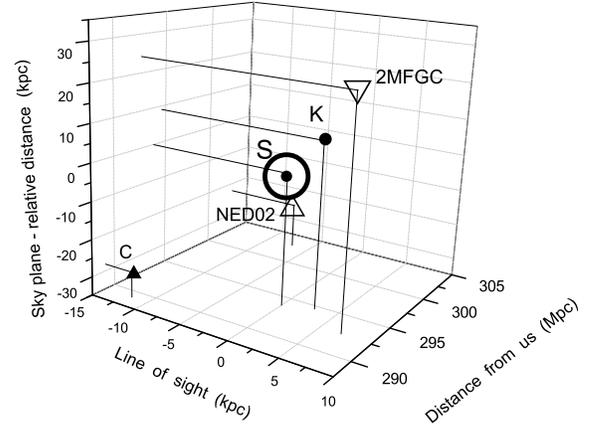}}
\caption{Spatial distribution of the main galaxies of the AM\,0546-324 system centered on the Shadowy galaxy: the line-of-sight in kpc (X-axis); the calculated distance from us in Mpc (Y-axis); and the relative distance between the objects in the sky-plane in kpc (Z-axis). The S galaxy is displayed as a dot inside a circle.}
\label{fig_10}
\end{figure}

\begin{figure}[h]
\centering
\resizebox{\hsize}{!}{\includegraphics[]{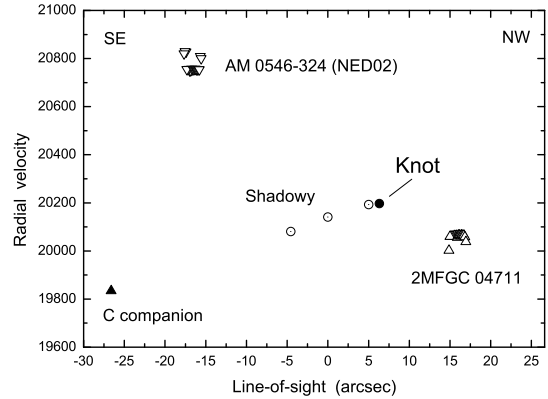}}
\caption{Calculated radial velocity in km\,s$^{-1}$ versus the distribution of the main objects in the line-of-sight centered on the Shadowy galaxy. The open circles are the center, the SE and NW sections of the S galaxy, respectively.}
\label{fig_11}
\end{figure}

The galaxy cluster \object{Abell S0546} has 23 cluster members between $m_{3}$ and $m_{3} + 2$ (Abell \cite{a1965}), and the whole \mbox{AM\,0546-324} system is the core of the S0546. There are quite a few dwarf satellites around it, as expected for a local gravitational potential well defined by the AM\,0546-324 system (see lower panel of Figs.~\ref{fig_02} and ~\ref{fig_09}). The main galaxies quoted in the literature are 2MFGC 04711, NED02, and the knot (K), which is described as an almost E1 galaxy (Quintana \& Ram\'{i}rez \cite{qr1995}). Other members quoted in our study are the S galaxy and the C companion, which have been confirmed as members of this cluster. The derived radial velocity of the S galaxy is 20\,141 km\,s$^{-1}$, which is our estimate of the radial velocity of the Abell S0546 core. The derived radial velocity of S using the non-relativistic formula agrees with the S0546 distance class $m_{10}=$ 5 (cz = 20\,893 km\,s$^{-1}$) extracted from Abell et al. (\cite{aco1989}). Figure~\ref{fig_10} displays the spatial distribution of the main galaxies of the AM\,0546-324 system centered on the S galaxy (see also Fig.~\ref{fig_11}).

Along the whole slit, the spectra of the three main galaxies show absorption lines characteristic of early-type objects. No star-forming regions and no nuclear ionization sources were detected. The whole AM\,0546-324 system seems to be tidally bound with radial velocity differences that range between 56$-$613 km\,s$^{-1}$. Adopting the S galaxy as the center of the system, the pair-velocity-difference combinations of the main objects are displayed in Table~\ref{table4}. The nearby C galaxy was partially inside the GMOS-S slit and the estimated redshift z=0.0685 corresponds to a heliocentric radial velocity of \mbox{$V$= 19\,834 $\pm$30 km\,s$^{-1}$}, with a calculated distance of 289.2 Mpc and a dynamical mass of 3.74$\times10^{10}$ M$_{\sun}$. The radial velocity of the K galaxy (20\,923 km\,s$^{-1}$) was extracted from Quintana \& Ram\'{i}rez (\cite{qr1995}). From the quoted mass, 2MFGC and NED02 each have approximately 31\% of the mass of the S galaxy, and the C galaxy has almost  7\%. Evidence of the tidal interaction of this system are seen in the external deformations of 2MFGC 04711, NED02, and K galaxies, as well as the rims in the S galaxy (displayed in Figs.~\ref{fig_01},~\ref{fig_02}, and~\ref{fig_03}).

In this section, we propose the following kinematical behavior for the AM\,0546-324 system based on spectroscopy and the direct image: (A) the S-galaxy is the center of the system, its SE-section is approaching and the NW-section is receding from us; (B) the 2MFGC 04711 galaxy  is approaching us, and seems to be embedded in the peripheral material of the S-galaxy, its motion is retrograde from de S-galaxy apparent rotation; (C) NED02 is receding from us and its motion is also retrograde from the S-galaxy apparent rotation; (D) the K-galaxy is receding from us, and seems to be embedded in S-galaxy material and coupled prograde with the S-galaxy motion; (E) the C-galaxy is approaching us and coupled prograde with the S-galaxy motion.

Both 2MFGC 04711 and NED02 show a U-shaped rotation profile. In addition to the argument that tidal coupling in ellipticals with no net rotation will result in a U-shaped rotation profile with the galaxy core at the base of the U (Borne \& Hoessel \cite{bh1988}; Borne et al. \cite{borne1994}), it was also suggested that when the galaxy has a low degree of internal rotation, the tidal coupling should produce this U-type profile (Borne et al. \cite{borne1994}). The lower panels of Figs.~\ref{fig_05} and~\ref{fig_06} show the rotation profile in the interval $\pm$1\arcsec \,for the 2MFGC 04711 and NED02, respectively. The oversampled data are displayed in both figures to show signatures around the kinematical center. These figures suggest that (1) 2MFGC 04711 seems to have a low degree of perturbation and there is maybe a very slow internal rotation, its kinematical center is almost centered at the U-shaped-base profile; (2) NED02 seems to show no internal rotation in its U-shaped profile, but there is a break indicated by the MgIb and H$\beta$ sampled data, for which we do not have a reliable explanation. The kinematical center is off-centered a few pc to the SE-direction in its U-shaped-base profile. This phenomenon is also seen in other pair interaction of Solitaire-type galaxies (see Fa\'{u}ndez-Abans et al. \cite{fa_oa2010}).

% ----------------------------> Table 4

\begin{table}
\caption{Radial velocity differences in the galaxy system.}
\label{table4}
\centering
\begin{tabular}{lc}
\hline \hline
Pair & $\bigtriangleup$V\,(km/s)\\
\hline
Shadowy$-$2MFGC 04711 & \hspace{+0.1cm}84 \\
Shadowy$-$Knot & \hspace{+0.1cm}56 \\
Shadowy$-$NED02 & 613 \\
Shadowy$-$C galaxy & 307 \\
2MFGC 04711$-$Knot &  140 \\
2MFGC 04711$-$NED02 & 697 \\
NED02$-$C galaxy & 920 \\
\hline
\end{tabular}
\end{table}

% ---------------------------------Figure 09

% ---------------------------------Figure 10

% ---------------------------------Figure 11

Figures~\ref{fig_02} (lower panel) and~\ref{fig_09} show some dwarf objects around the AM\,0546-324 system. A few of those objects may be aligned with the visible AM\,0546-324 structure. This sparse system has an intrinsic tidal field, which could be an interesting laboratory for studying the relationship between the central components and the dark matter halo in weak fields (see simulations for a denser cluster environment by Pereira \& Bryan \cite{pb2010}). A non-negligible anonymous galaxy is to the east between the field galaxy \object{2MASX J05481766-3239441} and the AM\,0546-324 system. The coordinates of the centroid (J2000), as calculated differentially from the centroid of 2MASX J05481766-3239441, are $\alpha  = $05$^{\rm h}$\,48$^{\rm m}$\,25\fm81 , \mbox{$\delta  =$ $-$32\degr\,39\arcmin\,49\farcs5}, with no previously reported redshift in the literature. This edge-on anonymous galaxy lies almost aligned with the linear distribution of the AM\,0546-324 system members. Is this object a candidate for tidal alignment? (see recent discussion on tidal alignment model of intrinsic galaxy alignments by Blazek et al. \cite{bmac2011})

The stellar formation history of 2MFGC 04711 and AM\,0546-324 (NED02) galaxies were well represented by the stellar population synthesis code {\sc STARLIGHT} (see Figs. \ref{fig_07} and \ref{fig_08}). The synthesis results 
in flux fraction as a function of the distance to the center of each galaxy do not show any spatial variation in the contribution of the different stellar population components. Both galaxies are dominated by an old stellar population with age  between $2\times10^{9} <\rm t \leq 13\times10^{9}$ yr in all apertures. 

\section{Conclusions}

We reported optical band spectroscopy observations of the AM\,0546-324 system, which is the core of Abell S0546 cluster of galaxies. Morphological substructures were found in an enhanced r-image of this system. This suggests that the members are presently undergoing early stages of tidal interaction. 

Below is a summary of our main results:

\begin{itemize}

\item The AM\,0546-324 system is composed of four main galaxies: 2MFGC 04711, AM\,0546-324 (NED02), the K galaxy, and the one named S galaxy by us. Adopting the S galaxy as the center of this gravitationally bound system, the radial velocity differences between the different quoted members vary from 43 to 646 km\,$s^{-1}$. 

\item Within 1.2 arcmin of AM\,0546-324 there are a few relevant field companions such as the C galaxy in the SE direction and a new Polar Ring galaxy candidate in the SW. Several dwarf objects in and surrounding this system are close enough to be candidate members of this system, but no quoted redshift for these objects was found in the literature.

\item The S galaxy seems to be large enough to wrap up all principal companions with its smooth distribution of material.  

\item The spectra of 2MFGC 04711, NED02, S, and the C galaxy resemble those of early-type galaxies and no emission lines were detected. No star-forming regions and no nuclear ionization sources were detected in the observed regions of the four main galaxies.

\item The calculated heliocentric radial velocity for the S galaxy is 20\,141 $\pm$ 10 km$s^{-1}$ ({\it z} = 0.0696), which agrees with the radial velocity of the Abell S0546 cluster \mbox{(cz = 20\,893 km\,s$^{-1}$)}; for 2MFGC 04711, it is 20\,057 $\pm$ 10 km$s^{-1}$ ({\it z} = 0.0693); and for NED02, it is 20\,754 $\pm$ 10 km$s^{-1}$ \mbox{({\it z} = 0.0718)},  both in agreement with quoted values in NED.

\item The C galaxy, cz = 19\,834 $\pm$ 40 km$s^{-1}$ ({\it z} = 0.0685), and the K galaxy, cz = 20\,197 ({\it z} = 0.0698), are both bound members of the AM\,0546-324 system.

\item From the calculated mass lower limit, both 2MFGC 04711 and NED02 have $\sim$31\% of the mass of the S galaxy, and the C galaxy, almost 7\%.

\item The rotation profiles of 2MFGC 04711 and NED02 are typical of tidal coupling in ellipticals with no net rotation, which results in a U-shaped rotation profile with the galaxy core at the base of the U. Both galaxies are  gravitationally coupled directly with the proposed central object of the cluster, the S galaxy. The  U-shaped structure is a direct observational signature of tidal friction with the extended material of the S galaxy. 

\item Internally, the no-net rotation core in the U-shaped rotation profile of both 2MFGC 04711 and NED02 seems to be slightly perturbed by the tidal interaction with the S galaxy, which lies in the center of the local gravitational potential-well of this system. 

\item  2MFGC 04711 and AM\,0546-324 (NED02) are completely dominated  by an old stellar population with age between $2\times10^{9} <\rm t \leq 13\times10^{9}$ yr.

\end{itemize}

In summary, AM\,0546-324 is a system where signatures of tidal perturbations and friction are clearly visible. The deformity detected in the 2MFGC 04711, NED02, and K galaxies is due to large tidal forces exerted principally by the S galaxy (like the deformation and dynamical friction between two elliptical galaxies$-$Prugniel \& Combes \cite{pc1992}). Simultaneously, the S galaxy is perturbed by the whole interaction with all principal objects of the system.

Two questions still remain to be answered: (1) is the S galaxy environment the starting point for the birth of a future cD galaxy? and (2) what is the origin of the S galaxy?  

\begin{acknowledgements}
      
This work was partially supported by the Ministerio da Ci\^{e}ncia, Tecnologia e Inova\,{c}\~{a}o
(MCTI), Laborat\'{o}rio Nacional de Astrof\'{i}sica, and Universidade do Vale do Para\'{i}ba - UNIVAP. A. C. Krabbe  thanks the support of FAPESP, process 2010/01490-3. We also thank Ms. Alene Alder-Rangel and M.~de Oliveira-Abans for editing the English in this manuscript. Based on observations obtained at the Gemini Observatory, which is operated by the Association of Universities for Research in Astronomy, Inc., under a cooperative agreement with the NSF on behalf of the Gemini partnership: the National Science Foundation (United States), the Science and Technology Facilities Council (United Kingdom), the National Research Council (Canada), CONICYT (Chile), the Australian Research Council (Australia), Ministerio da Ci\^{e}ncia, Tecnologia e Inova\,{c}\~{a}o (Brazil) and Ministerio de Ciencia, Tecnolog\'{i}a e Innovaci\'{o}n Productiva  (Argentina). The observations were performed under the identification number GS-2010B-Q-7. This publication makes use of data products from the Two Micron All Sky Survey, which is a joint project of the University of Massachusetts and the Infrared Processing and Analysis Center/California Institute of Technology, funded by the National Aeronautics and Space Administration and the National Science Foundation. 
This research also used the NASA/ IPAC Infrared Science Archive, which is operated by the Jet Propulsion Laboratory, California Institute of Technology, under contract with the National Aeronautics and Space Administration.

\end{acknowledgements}

\end{document}